\documentstyle[12pt,aps,epsfig,colordvi]{revtex}
\input{epsf}                                                               
\tightenlines
\begin{document}

\draft
                                                                           
\title{\bf {Observation of color-transparency in diffractive dissociation
of pions}}

%
%
\author{
    E.~M.~Aitala,$^9$
       S.~Amato,$^1$
    J.~C.~Anjos,$^1$
    J.~A.~Appel,$^5$
       D.~Ashery,$^{14}$
       S.~Banerjee,$^5$
       I.~Bediaga,$^1$
       G.~Blaylock,$^8$
    S.~B.~Bracker,$^{15}$
    P.~R.~Burchat,$^{13}$
    R.~A.~Burnstein,$^6$
       T.~Carter,$^5$
 H.~S.~Carvalho,$^{1}$
  N.~K.~Copty,$^{12}$
    L.~M.~Cremaldi,$^9$
 C.~Darling,$^{18}$
       K.~Denisenko,$^5$
       S.~Devmal,$^3$
       A.~Fernandez,$^{11}$
       G.~F.~Fox,$^{12}$
       P.~Gagnon,$^2$
       S.~Gerzon,$^{14}$
       C.~Gobel,$^1$
       K.~Gounder,$^9$
     A.~M.~Halling,$^5$
       G.~Herrera,$^4$
 G.~Hurvits,$^{14}$
       C.~James,$^5$
    P.~A.~Kasper,$^6$
       S.~Kwan,$^5$
    D.~C.~Langs,$^{12}$
       J.~Leslie,$^2$
       J.~Lichtenstadt,$^{14}$
       B.~Lundberg,$^5$
       S.~MayTal-Beck,$^{14}$
       B.~Meadows,$^3$
 J.~R.~T.~de~Mello~Neto,$^1$
    D.~Mihalcea,$^{7}$
    R.~H.~Milburn,$^{16}$
 J.~M.~de~Miranda,$^1$
       A.~Napier,$^{16}$
       A.~Nguyen,$^7$
  A.~B.~d'Oliveira,$^{3,11}$
       K.~O'Shaughnessy,$^2$
    K.~C.~Peng,$^6$
    L.~P.~Perera,$^3$
    M.~V.~Purohit,$^{12}$
       B.~Quinn,$^9$
       S.~Radeztsky,$^{17}$
       A.~Rafatian,$^9$
    N.~W.~Reay,$^7$
    J.~J.~Reidy,$^9$
    A.~C.~dos Reis,$^1$
    H.~A.~Rubin,$^6$
 D.~A.~Sanders,$^9$
 A.~K.~S.~Santha,$^3$
 A.~F.~S.~Santoro,$^1$
       A.~J.~Schwartz,$^{3}$
       M.~Sheaff,$^{17}$
    R.~A.~Sidwell,$^7$
    A.~J.~Slaughter,$^{18}$
    M.~D.~Sokoloff,$^3$
       J.~Solano,$^{1}$
       N.~R.~Stanton,$^7$
      R.~J.~Stefanski,$^5$ 
      K.~Stenson,$^{17}$
    D.~J.~Summers,$^9$
 S.~Takach,$^{18}$
       K.~Thorne,$^5$
    A.~K.~Tripathi,$^{7}$
       S.~Watanabe,$^{17}$
 R.~Weiss-Babai,$^{14}$
       J.~Wiener,$^{10}$
       N.~Witchey,$^7$
       E.~Wolin,$^{18}$
     S.~M.~Yang,$^{7}$
       D.~Yi,$^9$
       S.~Yoshida,$^{7}$
       R.~Zaliznyak,$^{13}$
       and
       C.~Zhang$^7$ \\
(Fermilab E791 Collaboration)
}

\address{
$^1$ Centro Brasileiro de Pesquisas F\'\i sicas, Rio de Janeiro, Brazil\\
$^2$ University of California, Santa Cruz, California 95064\\
$^3$ University of Cincinnati, Cincinnati, Ohio 45221\\
$^4$ CINVESTAV, Mexico\\
$^5$ Fermilab, Batavia, Illinois 60510\\
$^6$ Illinois Institute of Technology, Chicago, Illinois 60616\\
$^7$ Kansas State University, Manhattan, Kansas 66506\\
$^8$ University of Massachusetts, Amherst, Massachusetts 01003\\
$^9$ University of Mississippi-Oxford, University, Mississippi 38677\\
$^{10}$ Princeton University, Princeton, New Jersey 08544\\
$^{11}$ Universidad Autonoma de Puebla, Mexico\\
$^{12}$ University of South Carolina, Columbia, South Carolina 29208\\
$^{13}$ Stanford University, Stanford, California 94305\\
$^{14}$ Tel Aviv University, Tel Aviv, Israel\\
$^{15}$ Box 1290, Enderby, BC, VOE 1V0, Canada\\
$^{16}$ Tufts University, Medford, Massachusetts 02155\\
$^{17}$ University of Wisconsin, Madison, Wisconsin 53706\\
$^{18}$ Yale University, New Haven, Connecticut 06511\\
}

\date{\today}                                                        
\maketitle

\begin{abstract}

\Black{
We have studied the diffractive dissociation into di-jets of 500 
GeV/$c$ pions scattering coherently from carbon and platinum targets.
Extrapolating to asymptotically high energies (where $  t_{\rm min}
\, \to \, 0 $), we find that when the per-nucleus cross-section for this
process is parameterized as $ \sigma = \sigma_0 A^{\alpha} $,
$ \alpha $ has values near 1.6, the exact result depending on
jet transverse momentum.
These values are in agreement with those predicted
by theoretical calculations of color-transparency.}\\

\end{abstract}

 

\Black{
Color transparency (CT) is the name given to the prediction that
the color fields of QCD cancel for physically small singlet systems of
quarks and gluons \cite{lonus}.
This color neutrality (or color screening)
should lead to the suppression of initial and final state
interactions of small-sized systems
formed in hard processes \cite{ct},
often referred to as
point-like configurations (PLC's).
Observing
color transparency requires that a PLC is formed and that the energies are
high enough so that expansion of the PLC does not occur while
traversing the target \cite{ffs,jm,bm}
(the ``frozen" approximation).
To demonstrate that the interactions of a PLC with the nucleons
in a nucleus are suppressed compared to those of ordinary hadrons
requires identifying observables which depend explicitly on the
cross-sections of the PLC's.
Measurements of color transparency are important for clarifying the dynamics
of bound states in QCD \cite{fms1,fs}.\\
}

\Black{
Perturbative QCD predicts that a PLC
is formed 
in many two-body hadronic processes
at very large momentum transfer \cite{ct,liste}.
Experimental studies of such processes have failed to produce
convincing evidence of color transparency\cite{eva,slac,cebaf}.
However, the momentum transfer may not have been high enough and/or
the frozen approximation not valid under these experimental
conditions.
Evidence for color transparency 
(small hadronic cross-sections) has been observed in other 
types of processes:
in the A-dependence of
$J/\psi$ photoproduction \cite{e691}, 
in the $ Q^2 $-dependence of the $ t $-slope of diffractive $ \rho^0 $
production in muon scattering\cite{e665_1}
(where $ Q^2 $ is the invariant mass of the virtual photon and
$ t $ denotes the negative square of the 
momentum transfer from the virtual photon
to the target proton),
in the yield of non-diffractive $ \rho^0 $ production in 
deep inelastic muon scattering\cite{e665_2}, and in
the energy and flavor dependences of vector
meson production in $ ep $ scattering at HERA \cite{ha}.
In this paper, we report a direct observation of color transparency in the
$ A $-dependence of diffractive di-jet production by pions.\\
}

\Black{
The pion wave function can be expanded in terms of Fock states:
\begin{equation}
\Psi = \alpha |q\bar {q}\rangle + \beta |q\bar {q}g\rangle +
           \gamma |q\bar {q}gg\rangle + \cdots \, .
\end{equation}
The first (valence) component is dominant at large $Q^2$. The other
terms are suppressed by powers of $1/Q^2$ for each additional parton,
according to counting rules \cite{bf,stesto}.
When the relative velocities of all participating particles
are nearly light-like, time dilation lengthens the lifetimes
of these states and ``freezes" the partonic content of the pion
``seen" by the other particles.
Bertsch {\em et al.} \cite{bbgg} proposed that when high momentum pions hit
a nuclear target, the physically small  $|q\bar {q}\rangle$ components will be
filtered by the nucleus and materialize as
diffractive di-jets.
In a more recent calculation, based on a generalization of the QCD factorization
theorem, Frankurt {\em et al.} \cite{fms}  proposed that these small
$|q\bar {q}\rangle$ components can scatter coherently from
nuclei producing high mass di-jets. 
When the transverse momentum of the individual jets
with respect to the beam direction ($k_t$) is large,
the mass of the di-jet must also be large.
Frankurt {\em et al.} show that for $k_t > 1.5 \, {\rm GeV}/c$ 
the interaction with
the nucleus is completely coherent and $ \sigma ( |q\bar {q}\rangle
{\cal N}  \to  $di-jets $ \cal{N} $) is small. 
This leads to  an
$ A^2 $ dependence of the forward amplitude squared for asymptotically high
energies (where $ t_{\rm min} \to 0 $).\\
}

\Black{
As a good approximation, the integrated
{\em per-nucleus} cross-section
for producing these high mass di-jets grows
as $ A^{4/3} $.
The forward amplitude squared provides a factor of $ A^2 $ \cite{fms}
and the
integral of the elastic-scattering form factor,
$ \approx \int exp ( - \beta R_0^2  t ) \, dt $,
contributes a factor $ A ^ {-2/3} $ (on the assumption
that $ R_0  \approx A^{1/3} $  and $ \beta $, which depends on nuclear
density, is the same for all targets).
This should be compared with $ \sigma \propto A^ {2/3} $
typical of normal pion-nucleus interactions (for which
the shadowing cross-section is approximately the full $ \pi  $-$ \cal{N} $
inelastic cross-section).
Using more realistic nuclear form factors and accounting for
higher twist effects will change the $ A $-dependence modestly, and
electromagnetic contributions should produce negligible
effects\cite{fms00}.\\
}

\Black{
In this Letter we report measurements of the 
$ A $-dependence of the diffractive dissociation
into di-jets of 500 GeV/$c$ pions 
scattering coherently from carbon and platinum targets using data
from Fermilab experiment E791.
We recorded $2\times10^{10}$  $\pi^-$-nucleus
interactions  during the 1991/92 fixed-target run at Fermilab using an open
geometry spectrometer\cite{e791} in the Tagged Photon Laboratory. 
The segmented  target
consisted of one platinum foil and four diamond foils separated by gaps of
1.34 to 1.39 cm. 
Each foil was approximately 0.4\% of an interaction length
thick (0.5 mm for platinum  and 1.6 mm  for carbon). 
Six planes
of silicon microstrip detectors (SMD) and eight proportional wire  chambers
(PWC)  were used to track the beam particles. 
The downstream detector
consisted of 17 planes of SMDs for vertex detection, 35 drift chamber planes,
two PWCs, two magnets for momentum analysis, two multi-cell threshold
\v{C}erenkov counters
(not used in this analysis),
electromagnetic and hadronic calorimeters for electron identification
and
for  online triggering, and two planes of muon scintillators.
An interaction trigger required a beam particle and
an interaction in the target. 
A very loose  transverse energy trigger, based
on the energy deposited in the calorimeters, and a fast data acquisition
system allowed us to collect data at a rate of 30 Mbytes/s with
50$\mu$s/event dead time and to write data to tape at a rate of 10 Mbytes/s.\\
} 

\Black{
Data reconstruction and additional event selection were done using offline
parallel processing systems.
The data for this analysis are selected with the primary
requirement that at least 90\% of the beam
momentum is carried by charged particles. 
This reduces the effects of unobserved neutral particles.
In addition, this analysis uses events produced relatively early 
in the experiment, before 
the performance of the drift chambers
degraded in the region through which the pion beam passed.
The offline selection for this analysis was implemented
after most of the data had already been filtered for other analyses,
and we use data taken only when all five targets were
in place.
In all, about 10\% of the experiment's integrated data set is used in this 
analysis.\\
} 

\Black{
The JADE  jet-finding algorithm \cite{jade} is used to identify
two-jet events.
The algorithm's cut-off parameter ($m_{cut}$) was optimized for 
this analysis using Monte Carlo simulations which are described
below.
For each two-jet event
we calculate the transverse momentum of each jet with
respect to the beam axis ($ k_t $), the transverse
momentum of the di-jet system with respect to the beam axis
($ q_t $), and the di-jet invariant mass, $ M_J $.
The di-jet invariant mass is related to the quarks'
longitudinal momentum fractions in the pion
infinite momentum frame ($ x $) by simple kinematics:
$ M_J^2 = {k_t^2} / [x ( 1-x ) ] $.
To assure clean selection of high-mass di-jet
events, a minimum $k_t$ of 1.2 GeV/$c$ is required. The selection of clean
di-jet events was verified by testing their relative azimuthal angle which
for pure di-jets should be near $180^{\circ}$. This angle is required to
lie within $20^{\circ}$ of this value.
The size of a $ | q \bar q \rangle $ system which produces di-jets
with $ k_t > $ 1.5 GeV/$c$ can be estimated as $ 1 / Q  \le 0.1 \, {\rm fm}$ 
where 
$ Q^2 \sim M_J^2 \ge 4 k_t^2 \sim 10 \ {\rm GeV}^2 / c^2 $.
The distance that the $ | q \bar q \rangle $ system travels before it
expands appreciably, the coherence length, is given by 
$ \ell_c \sim ( 2p_{\rm lab} ) / ( M_J^2 - m_\pi^2) $ \cite{ffs} which is
$ \sim 10 $ fm for $ M_J \sim 5 $ GeV/$c^2$.
Therefore, we expect that the di-jet signal events selected in this
analysis evolve from point-like configurations which will exhibit
color transparency. 
} 

\begin{figure}[h]
\centerline{\epsfxsize=15cm \epsfbox{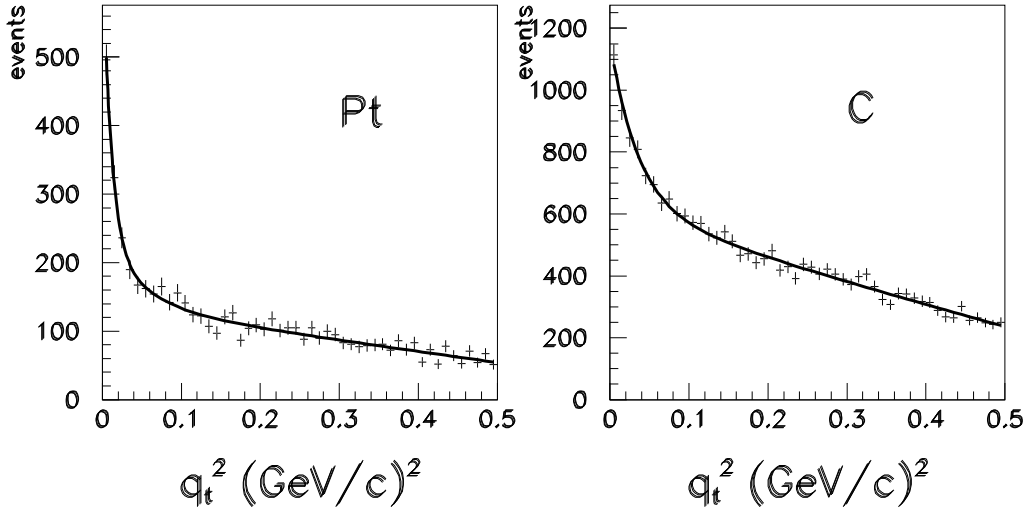}}
\vglue -6.0cm
\caption{$q_t^2$ distributions of di-jets with $k_t ~\geq 1.25
~\rm{GeV/c}$ 
from interactions of 500 GeV/c
$\pi^-$ with platinum and carbon targets.}
\label{data_diff}
\end{figure}

\Black{
The $ q_t^2 $ distributions of the selected di-jet events are shown
in Fig.~\ref{data_diff}.
The peaks at small $ q_t ^2$ arising from coherent scattering from
nuclei are smeared due to missing neutrals and detector resolution,
but the integrated coherent
signals can be extracted with reasonable accuracy.
Before detector acceptance and smearing, 
coherent peaks should be produced with\
$ dN/d q_t^2 \, \propto \, {\rm exp}(-b q_t^2) $ with  $ b $
inversely proportional to the nucleus's radius 
(2.44 fm for carbon and 5.27 fm for platinum\cite{rad}).
Because theory predicts that the $ A $-dependence varies with $ k_t $
\cite{fms}, the
analysis is carried out in three $k_t$ regions:
$1.25 \ \rm{GeV/}c ~\leq ~k_t ~\leq ~1.5 \ \rm{GeV/}c$,
$1.5 \ \rm{GeV}/c ~ < ~k_t ~\leq ~2.0 \ \rm{GeV}/c$, and
$2.0 \ \rm{GeV/}c ~ < ~k_t ~\leq ~2.5 \ \rm{GeV/}c$.
Altogether, we find about 5000 coherent di-jet events in the
carbon data set and about 2800 in the platinum data set.\\
} 

\Black{
To determine the relative number of coherent events produced  in
each target,
we fit the data of Fig. \ref{data_diff}
as sums of  $ q_t^2 $
distributions of di-jet events produced coherently
from nuclear targets, of di-jet events produced coherently
from individual nucleons but incoherently with
respect to the nuclear targets,
and of background.
The shapes of the di-jet distributions are calculated using
Monte Carlo simulations.
We use the LUND PYTHIA-JETSET package \cite{mc}
to generate di-jet events with masses of  4, 5, and 6 GeV/c$^2$.
This covers the range of $ k_t $ observed in the data.
The quark momentum distribution inside the pions is generated using an
asymptotic wave function \cite{lb,er} which is consistent with
the data presented in our companion paper\cite{wf_prl}.
Coherent nuclear events are generated with $ q_t^2 $ slopes 
appropriate to carbon and platinum.
Coherent nucleon events are generated with $ q_t^2 $ slopes
appropriate to the nucleon radius (0.8 fm), truncated at
$ q_t^2 = 0.015 $ $ ( {\rm GeV}/c )^2 $ to account for
the nucleon binding energy.
The generated events are passed through a 
detector simulation and digitized to mimic real events.
They are reconstructed and analyzed using the same
programs used to reconstruct and analyze the real data.\\
} 

\begin{figure}[b]
\centerline{\epsfxsize=8.0cm \epsfbox{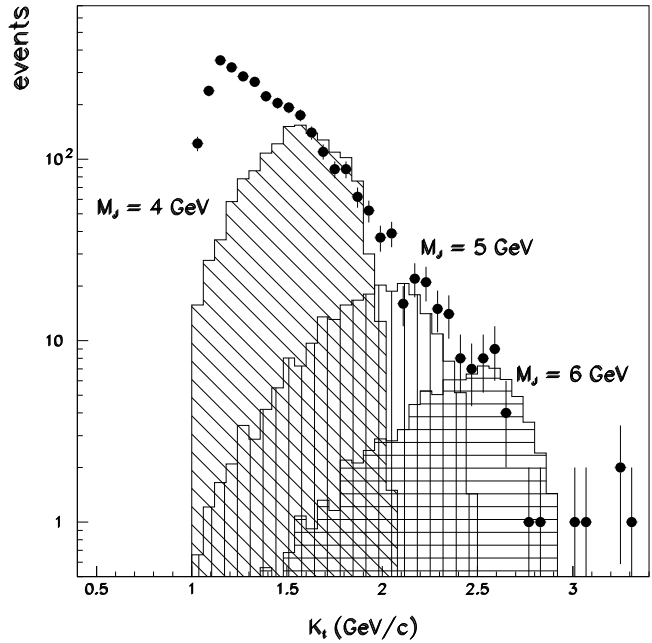}}
\caption{$k_t$ distributions of simulated di-jets with 4 (slanted lines),
5 (vertical lines), and 6 GeV/c$^2$ (horizontal lines) masses normalized
and superimposed on the distribution from data taken with a platinum 
target.}
\label{kt}
\end{figure}

\begin{figure}
\centerline{\epsfxsize=16.0cm 
\epsfbox{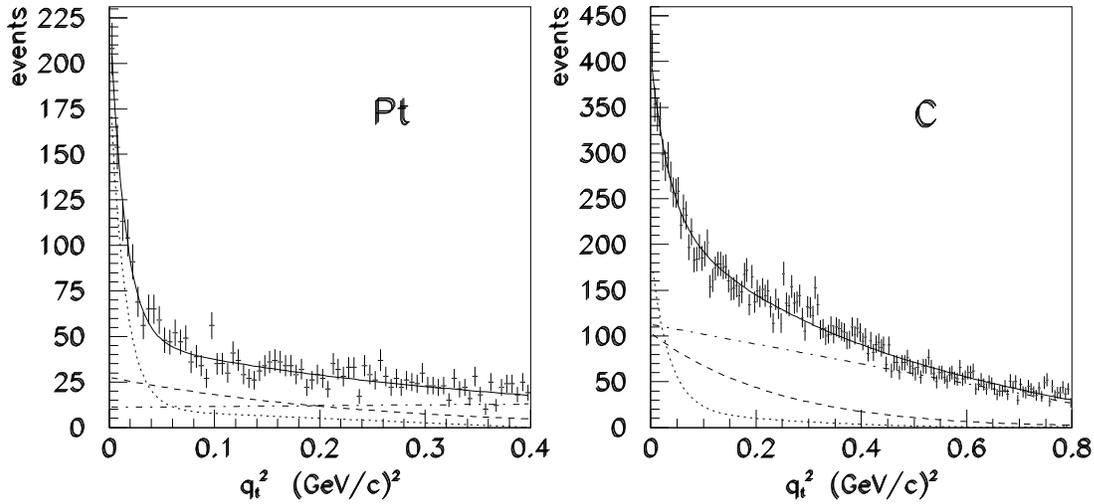}}
\vglue -7.0cm
\caption{$q_t^2$ distributions of di-jets with $1.5 \leq k_t \leq 2.0$
GeV/c for the platinum and carbon targets. The lines are fits of the MC
simulations to the data: coherent nuclear dissociation (dotted line), 
coherent nucleon/incoherent nuclear
dissociation (dashed line), background (dashed-dotted line) and total fit
(solid line).}
\label{difr_a}
\end{figure}

\Black{
To determine the relative efficiencies for observing di-jet events
produced through coherent diffractive scattering from carbon and
platinum nuclei, we use Monte Carlo samples generated with
di-jet masses of 4, 5, and 6 GeV/$c^2$.
The proportions are adjusted to reproduce the $ k_t $ spectrum
of the data, as observed in Fig. \ref{kt}.
For each $ k_t $ range, and for each target, 
we fit the $ q_t^2 $ distribution using the
signal shapes from the Monte Carlo simulations
and assuming the background contribution is linear in $ q_t ^2 $.
Fig.
\ref{difr_a} shows the results of the fit for
$1.5 \ \rm{GeV}/c ~\leq  ~k_t ~\leq ~2.0 \ \rm{GeV}/c$. 
The dashed
line shows the coherent nuclear distribution, 
the dotted line the coherent nucleon/incoherent nuclear
distribution,
and the dash-dotted line the residual background. 
This
background represents the components of the data which are not
simulated well, such as badly identified jets. 
The background's contribution in the region of
the coherent distribution is small. 
These fits provide
normalization factors between the number of simulated events of each
kind and the data.\\
} 

\Black{
We derive the relative numbers of produced di-jet events 
for each target in each $k_t$ bin by integrating over the diffractive terms
in the fits and accounting for the relative efficiencies as described
above. 
The signals from the carbon and platinum targets in any one $ k_t $
range have slightly different mass distributions, and we correct
the relative yields to account for this.
Using the
measured target thicknesses, we determine the ratio of cross-sections for
coherent production of diffractive di-jets from the carbon and
platinum targets
(which received essentially the same beam flux).
Theoretical calculations of color transparency at asymptotically high energies
predict $ \sigma = \sigma_0 A^{\alpha} $
with $ \alpha $ = 4/3
assuming very simple nuclear wave
functions.
At the energy of this experiment,
$ | t_{\rm min} | > 0 $ reduces the cross-sections for 
coherent scattering on carbon (platinum) targets to 0.98 (0.93), 
0.97 (0.87), and 0.94 (0.76) times their asymptotic
high-energy values for $ M_J $ = 4.2, 5.0, and 6.0 GeV/$c^2$
(masses relevant for our $ k_t^2 $ bins).
We extrapolate our calculations of
$ \alpha $ to asymptotically high energies dividing the yields by
these factors.
The results for each $ k_t $ bin
are listed in Table \ref{alpha}.
Using more realistic wave functions, the predicted value of
the asymptotic value of $\alpha$ is 1.45 for carbon and platinum
targets. 
Frankfurt et al. \cite{fms} predict some dependence
of $\alpha$ on $k_t$ as well.
These values,
labelled $\alpha$(CT), are also listed in Table \ref{alpha}.\\
} 

\Black{
We have considered the sources of systematic
uncertainty which are listed in Table~\ref{tab:sys}.
The degree to which the simulations represented correctly the effect of
not including the neutral component of the jets is checked by raising the
minimum total momentum of charged particles from  450 GeV/c to 470 GeV/c.
The difference in the final results of $\alpha$ with and without this
requirement is taken to be the corresponding systematic uncertainty 
(``effect of neutrals").
The uncertainty due to using discrete masses in the Monte Carlo
simulation is estimated using the difference 
between results assuming that all the events in a given $k_t$
range have one mass or another (``discrete masses"). 
A third uncertainty is assigned to the change in yields due to
mass-distribution differences in carbon and platinum. 
We also observe 
some sensitivity to the fitting range used;
the associated
differences are taken as a fourth systematic uncertainty.
The total systematic
uncertainty is taken by adding these contributions in quadrature,
retaining the signs when not symmetric.\\
} 

\Black{
In summary, we have measured the relative cross-sections for diffractive
dissociation into di-jets of 500 GeV/$c$
pions scattering from carbon and platinum targets.
Extrapolating to asymptotically high energies (where
$ t_{\rm min} \to 0 $), we find that when the cross-section
is parameterized as $ \sigma = \sigma_0 A^{\alpha} $,
$ \alpha \sim 1.6 $.
This is consistent with expectations based on calculations of
color-transparency models and is clearly inconsistent with
the $\sigma \propto A^{2/3}$ dependence observed for inclusive
$ \pi $-nucleus scattering.
We have measured $ \alpha $ in three ranges of $ k_t $;
because the uncertainties are large, the results are consistent
with no variation, but also with the predicted variation.
The clear diffractive structure of the signals and variation of
the coherent cross-section with $ A $ indicate that we have observed
the coherent scattering of 
$ | q \bar q \rangle $ point-like configurations predicted
by color-transparency.\\
} 

We thank Drs. S.J. Brodsky, L. Frankfurt, G.A. Miller, and M. Strikman
for many fruitful discussions.
We gratefully acknowledge the staffs of Fermilab and of all the
participating institutions. This research was supported by the Brazilian
Conselho Nacional de Desenvolvimento Cient\'\i fico e Technol\'ogio,
the Mexican Consejo Nacional de Ciencia y Tecnologica,
the Israeli Academy of Sciences and Humanities, the
United States Department
of Energy, the U.S.-Israel Binational Science Foundation, and the United
States National Science Foundation. Fermilab is operated by the
Universities
Research Association, Inc., under contract with the United States
Department of Energy.

\begin{table}[H]
\caption{The exponent in $\sigma \propto A^{\alpha}$, experimental
results for coherent dissociation and the Color-Transparency
(CT) predictions.} 
\begin{tabular}{|c|c|c|c|c|c|}
$k_t$ bin & $\alpha$ & $\Delta \alpha_{stat}$ &$\Delta \alpha_{sys}$ & 
$\Delta \alpha$ & $\alpha$ (CT) \\
GeV/c & & & & & \\ \hline
1.25 $-$ 1.5 & 1.64     & $\pm$0.05  & +0.04 $-$0.11 &+0.06 $-$0.12
&1.25 \\
1.5 $-$ 2.0 & 1.52     & $\pm$0.09 & $\pm$0.08 &$\pm$0.12  & 1.45 \\
2.0 $-$ 2.5 & 1.55 & $\pm$0.11 & $\pm$0.12 &$\pm$0.16  & 1.60 \\
\end{tabular}
\label{alpha}
\end{table}

\begin{table}[H]
  \begin{center}
    \caption{The systematic errors} 
    \leavevmode  
    \begin{tabular}{|c|c|c|c|c|c|}
$k_t$ bin& effect of &discreet & different efficiency & fit range
& total \\ 
 GeV/c  & neutrals  & masses  & for C and Pt & sensitivity
&  \\ \hline
1.25 $-$ 1.5 & $-$0.09 &$+$0.03 $-$0.06  & $\pm$0.02 & $\pm$0.02 &$+$0.04
$-$0.11 \\
1.5 $-$ 2.0 & $\pm$0.03 & $\pm$0.03 & $\pm$0.06 & $\pm$0.04 &$\pm$0.08\\
2.0 $-$ 2.5 & $\pm$0.06 & $\pm$0.05 & $\pm$0.06 & $\pm$0.07 &$\pm$0.12 \\
    \end{tabular}
\vglue 0.5cm
    \label{tab:sys}
  \end{center}
\end{table}



\end{document}